\documentclass[aps, nofootinbib, superscriptaddress]{revtex4-2}

\usepackage{amsmath}
\usepackage{amssymb}
\usepackage{graphicx}
\usepackage[utf8]{inputenc}

\usepackage{color}

\newcommand{\hide}[1]{}

\begin{document}

\title{Saturation of Pauli blocking in near-extremal charged Nariai black holes}

\author{Chiang-Mei Chen} \email{cmchen@phy.ncu.edu.tw}
\affiliation{Department of Physics, National Central University, Chungli 32001, Taiwan}
\affiliation{Center for High Energy and High Field Physics (CHiP), National Central University, Chungli 32001, Taiwan}
\affiliation{Asia Pacific Center for Theoretical Physics, Pohang 37673, Korea}

\author{Chun-Chih Huang} \email{makedate0809@gmail.com}
\affiliation{Department of Physics, National Central University, Chungli 32001, Taiwan}

\author{Sang Pyo Kim} \email{sangkim@kunsan.ac.kr}
\affiliation{Department of Physics, School of Science and Technology, Kunsan National University, Kunsan 54150, Korea}
\affiliation{Center for High Energy and High Field Physics (CHiP), National Central University, Chungli 32001, Taiwan}
\affiliation{Asia Pacific Center for Theoretical Physics, Pohang 37673, Korea}
\date{\today}

\begin{abstract}
We solve the Dirac equation for a massive charged fermion in the near-extremal charged Nariai black hole with the near-horizon geometry $\mathrm{dS}_2 \times \mathrm{S}^2$.
At the one-loop level, contrary to the catastrophic emission of charged spinless bosons in [C.-M. Chen \textit{et al.}, Phys. Rev. D 110, 085020 (2024)], we show that the emission of fermions in a narrow time-like inner region between the cosmological horizon and black hole horizon saturates the bound from the Pauli blocking and does not give an amplification (quantum superradiance). Using the reciprocal relation, we find the Schwinger emission of fermions from $\mathrm{AdS}_2 \times \mathrm{S}^2$, and compare the Schwinger emission of fermions and bosons from near-extremal Nariai black holes.
\end{abstract}

\maketitle

\section{Introduction}

Extremal or near-extremal black holes have the enhanced symmetry~\cite{Bardeen:1999px, Kunduri:2007vf}. The near-extremal Reissner-Nordstr\"{o}m (RN) black holes in an asymptotically flat space or anti-de Sitter (AdS) space have the near-horizon geometry of ${\rm AdS}_2 \times {\rm S}^2$, whereas RN black holes in a de Sitter (dS) space with three horizons have two coincident limits: one is the coincidence of inner (Cauchy) and outer (event) horizon of black hole with the near-horizon geometry of ${\rm AdS}_2 \times {\rm S}^2$ and the other is the coincidence of black hole horizon and cosmological horizon, the so-called Nariai limit, with the geometry ${\rm dS}_2 \times {\rm S}^2$.

The low temperature quantum theory of charged black holes with $\mathrm{AdS}_2$ horizons provides an effective theory of the Einstein-Maxwell theory~\cite{Sachdev:2019bjn}. Recently the enhanced symmetries have been used to study quantum corrections of the near-extremal black holes and their entropies~\cite{Almheiri:2016fws, Nayak:2018qej, Iliesiu:2020qvm, Banerjee:2023quv, Turiaci:2023wrh, Banerjee:2023gll, Maulik:2025phe, Blacker:2025zca}.
It is argued that two-dimensional Jackiw-Teitelboim (JT) gravity is an appropriate effective field theory that dominates the near-extremal black holes at low temperature, and furthermore, this description may go away from the horizon~\cite{Kolanowski:2024zrq, Castro:2025itb}.

The authors have studied Schwinger emission of charged scalars (spin-0 bosons) from near-extremal charged black holes, in which the extremely low Hawking temperature suppresses Hawking radiation of charged particles as well as charge neutral particles while the electric field on the horizon provides a channel for charge emission via Schwinger mechanism~\cite{Schwinger:1951nm}. In fact, the enhanced symmetry allows one to solve the Klein-Gordon equation for massive charged scalars in near-extremal RN black holes~\cite{Chen:2012zn, Chen:2020mqs, Cai:2020trh, Chen:2023swn} and in near-extremal Kerr-Newman black holes~\cite{Chen:2016caa, Chen:2017mnm, Chen:2021jwy, Chen:2024ctf}, both in terms of hypergeometric functions. Recently we have found the Schwinger effect of spontaneous pair production of massive spinors (spin-1/2 fermions) by a uniform electric field in the ${\rm (A)dS}_2$~\cite{Chen:2025xrv}.

In this paper we will study the spontaneous production of spin-1/2 massive charged fermions from near-extremal charged Nariai black holes.
RN black holes in a dS space have three horizons: the inner (Cuachy) horizon, the outer (event) horizon and the cosmological horizon.
Near-extremal charged Nariai black holes have the black hole horizon close to the cosmological horizon, in which the black hole horizon emits Hawking radiation with Hawking temperature and the cosmological horizon emits with Gibbons-Hawking temperature. Though they are extremely small with slightly higher Hawking temperature than the Gibbons-Hawking temperature, the electric field that threads out from the black hole forces the black hole horizon to emit charges with the same sign as the black hole and the cosmological horizon to emit charges of the opposite sign (for the scalar case, see~\cite{Chen:2023swn, Chen:2024ctf}).
We solve the Dirac equation in the near-extremal charged Nariai black holes, find the Bogoliubov coefficients and therefrom the mean number of spontaneously produced fermions in the inner region and outer region. We then show that two narrow horizons produce fermion pairs still obeying the Pauli exclusion principle and saturating the Pauli blocking when two horizons coincide.

The organization of this paper is as follows. In Sec.~\ref{outer} we solve the Dirac equation for spin-1/2 massive charged fermion in the spacelike outer region of near-extremal Nariai black holes, find the Bogoliubov transformation and relation, and compute the mean number of fermion pair production. In Sec.~\ref{inner} we find the Bogoliubov transformation and relation, and then compute the mean number of fermion pair production in the timelike inner region of near-extremal Nariai black holes. In Sec.~\ref{(A)dS_2} we compare the fermion production in an exact Nariai black hole and in a uniform electric field in $\mathrm{dS}_2$ space, and apply the reciprocal relation of mean numbers to find the fermion production in ${\rm AdS}_2$ space. In Sec.~\ref{spinor_scalar} we compare the fermion emission with catastrophic scalar emission from near-extremal Nariai black holes.

\section{Outer Region of Near-Extremal Nariai Black Holes} \label{outer}

\subsection{Dirac Equation}
The Dirac equation for a massive charged spinor field, $\Psi$, in a charged black hole is given by
\begin{eqnarray}
\gamma^a e_a{}^\mu \left( \partial_\mu + \Gamma_\mu - i q A_\mu \right) \Psi + m \Psi = 0,
\end{eqnarray}
where $m$ and $q$ are the mass and charge of the spinor field, respectively, and $A_\mu$ is the gauge potential of black hole.
The tetrad $e_a{}^\mu$, one-form $\Gamma$, and connection one-form $\omega_{ab}$ are defined as
\begin{eqnarray}
\eta_{ab} = e_a{}^\mu e_b{}^\nu \, g_{\mu\nu}, \qquad
\Gamma = \Gamma_\mu dx^\mu = \frac{1}{4} \gamma^a \gamma^b \omega_{ab}, \qquad
d\vartheta^a = - \omega^a{}_b \wedge \vartheta^b,
\end{eqnarray}
where the gamma matrices satisfy $\gamma^a \gamma^b + \gamma^b \gamma^a = 2 \eta^{ab}$.
Below the Greek alphabet will denote coordinate indices, and the Latin alphabet denote frame indices.

We first consider the Dirac equation in the spacelike outer region of the near-extremal Nariai black hole, which is described by the geometry of ${\rm dS_2 \times S^2}$~\cite{Chen:2023swn}
\begin{eqnarray} \label{Nariai-metric}
ds^2 = r_\mathrm{ds}^2 \left( - \frac{d \tau^2}{f(\tau)} + f(\tau) d\rho^2 \right) + r_n^2 \left( d\theta^2 + \sin^2\theta d\varphi^2 \right), \qquad
f(\tau) = \tau^2 - B^2,
\end{eqnarray}
and the gauge (vector) potential is given by
\begin{eqnarray}
A = \frac{r_\mathrm{ds}^2 Q_n}{r_n^2} \tau d\rho = \phi(\tau) d\rho.
\end{eqnarray}
In the orthonormal frame $\vartheta^a = e^a{}_\mu dx^\mu$, we define the tetrad as
\begin{eqnarray}
e^0{}_\tau = \frac{r_\mathrm{ds}}{\sqrt{f}}, \qquad
e^1{}_\rho = r_\mathrm{ds} \sqrt{f}, \qquad
e^2{}_\theta = r_n, \qquad
e^3{}_\varphi = r_n \sin\theta, \label{out_tetrad}
\end{eqnarray}
and then write the term $\gamma^a e_a{}^\mu \Gamma_\mu$ as
\begin{eqnarray}
\gamma^a e_a{}^\mu \Gamma_\mu = \gamma^1 e_1{}^\rho \Gamma_\rho + \gamma^3 e_3{}^\varphi \Gamma_\varphi = \frac{\gamma^0}{4 r_\mathrm{ds} \sqrt{f}} \frac{df}{d\tau} + \frac{\gamma^2}{2 r_n} \cot\theta.
\end{eqnarray}

Setting the spinor field as
\begin{eqnarray}
\Psi = f^{-1/4} (\sin\theta)^{-1/2} \, \Phi, \label{Phi_spinor}
\end{eqnarray}
the Dirac equation can be written as
\begin{eqnarray} \label{Dirac_outer}
\left( \gamma^0 \frac{\sqrt{f}}{r_\mathrm{ds}} \partial_\tau + \frac{\gamma^1}{r_\mathrm{ds} \sqrt{f}} (\partial_\rho - i q \phi) + \frac{\gamma^2}{r_n} \partial_\theta + \frac{\gamma^3}{r_n \sin\theta} \partial_\varphi + m \right) \Phi = 0.
\end{eqnarray}

\subsection{Exact Solution}
We expand the spinor field~\eqref{Phi_spinor} in terms of the orthonormal spherical spinors
\begin{eqnarray}
\Phi^+_{\upsilon, n}(\theta, \varphi) = \begin{bmatrix} i \mathbb{Y}^n_{j \mp 1/2}(\theta, \varphi) \\ 0 \end{bmatrix}, \qquad \Phi^-_{\upsilon, n}(\theta, \varphi) = \begin{bmatrix} 0 \\ \mathbb{Y}^n_{j \pm 1/2}(\theta, \varphi) \end{bmatrix},
\end{eqnarray}
where each entry is a two-component spherical spinor defined in terms of the spherical harmonics.
The parameter
\begin{eqnarray}
\upsilon = \mp \left( j + \frac{1}{2} \right) \quad \textrm{for} \quad  j = l \pm \frac{1}{2},
\end{eqnarray}
is determined by the quantum number of the angular momentum $j$ and its projection $n$, where $-l \le n \le l $ and $l$ is an integer.
Moreover, the spherical spinors satisfy the following properties
\begin{eqnarray}
\gamma^0 \Phi^\pm_{\upsilon, n} = \mp i \Phi^\pm_{\upsilon, n}, \qquad
\gamma^1 \Phi^\pm_{\upsilon, n} = \pm i \Phi^\mp_{\upsilon, n}, \qquad
\left( \gamma^2 \partial_\theta + \gamma^3 \frac{\partial_\varphi}{\sin\theta} \right) \Phi^\pm_{\upsilon, n} = i \upsilon \Phi^\mp_{\upsilon, n}.
\end{eqnarray}
Now, we assume that the solution of the spinor field $\Phi$ is given by
\begin{eqnarray}
\Phi(\tau, \rho, \theta, \varphi) = \xi_+(\tau) \exp(i k \rho) \begin{bmatrix} i \mathbb{Y}^n_{j - 1/2}(\theta, \varphi) \\ 0 \end{bmatrix}
+ \xi_-(\tau) \exp(i k \rho) \begin{bmatrix} 0 \\ \mathbb{Y}^n_{j + 1/2}(\theta, \varphi) \end{bmatrix}.
\end{eqnarray}
Hereafter we will consider only $\upsilon = - (j + 1/2)$.

Then the Dirac equation can be separated into a pair of coupled equations
\begin{eqnarray} \label{eq_out_xi}
\left( \sqrt{f} \frac{d}{d\tau} \pm i m r_\mathrm{ds} \right) \xi_\pm + \left( \frac{i k - i q \phi}{\sqrt{f}} \mp \upsilon \frac{r_\mathrm{ds}}{r_n} \right) \xi_\mp = 0.
\end{eqnarray}
To solve Eq.~\eqref{eq_out_xi}, we first introduce new functions
\begin{eqnarray}
T_\pm(\tau) = \xi_+(\tau) \pm \xi_-(\tau) \quad \Rightarrow \quad \xi_\pm(\tau) = \frac{T_+(\tau) \pm T_-(\tau)}{2},
\end{eqnarray}
and then obtain
\begin{eqnarray} \label{eq_out_T}
\left( \sqrt{f} \frac{d}{d\tau} \pm \frac{i k - i q \phi}{\sqrt{f}} \right) T_\pm + \left( i m r_\mathrm{ds} \pm \upsilon \frac{r_\mathrm{ds}}{r_n} \right) T_\mp = 0.
\end{eqnarray}
Note that the function $f$ have been combined in Eq.~\eqref{eq_out_T}.

To further simplify Eq.~\eqref{eq_out_T}, we introduce a new variable and apply a transformation to the functions $T_\pm$ as
\begin{eqnarray}
z = \frac{\tau + B}{2 B}, \qquad
T_\pm(z) = \Sigma^{\mp 1} \tilde{T}_\pm(z), \qquad
\Sigma = z^{- i (\tilde{\kappa} + \kappa)/2} (z - 1)^{i (\tilde{\kappa} - \kappa)/2},
\end{eqnarray}
where $\tilde{\kappa} = k/B$ and $\kappa = r_\mathrm{ds}^2 (q Q_n/r_n^2) = r_\mathrm{ds}^2 E_n$ is the dimensionless electric field on the horizon.
Substituting these into Eq.~\eqref{eq_out_T}, we obtain
\begin{eqnarray} \label{eq_out_tT}
\Sigma^{\mp 2} \sqrt{z (z - 1)} \, \frac{d \tilde{T}_\pm}{dz} + \left( i m r_\mathrm{ds} \pm \upsilon \frac{r_\mathrm{ds}}{r_n} \right) \tilde{T}_\mp = 0.
\end{eqnarray}
Finally, after decoupling $\tilde{T}_\pm$, we find the solutions of Eq.~\eqref{eq_out_tT} in terms of the hypergeometric functions
\begin{eqnarray}
\tilde{T}_\pm &=& C_\pm F\left( \pm i (\mu + \kappa), \mp i (\mu - \kappa); \frac{1}{2} \mp i (\tilde{\kappa} - \kappa); 1 - z \right)
\nonumber\\
&+& D_\pm \Sigma^{\pm 2} \sqrt{z (z - 1)} \, F\left( 1 \mp i (\mu + \kappa), 1 \pm i (\mu - \kappa); \frac{3}{2} \pm i (\tilde{\kappa} - \kappa); 1 - z \right),
\end{eqnarray}
where $\mu^2 = \kappa^2 + m^2 r_\mathrm{ds}^2 + \upsilon^2 r_\mathrm{ds}^2/r_n^2$.
The relations of coefficients $C_\pm$ and $D_\pm$ can be determined by Eq.~\eqref{eq_out_tT}
\begin{eqnarray}
D_\pm = - C_\mp \frac{i m r_\mathrm{ds} \pm \upsilon r_\mathrm{ds}/r_n}{1/2 \pm i (\tilde{\kappa} - \kappa)}.
\end{eqnarray}

\subsection{Asymptotic and Near Horizon Behaviors}
Now we find the Bogoliubov relation between the modes at $\tau \rightarrow B$ and $\tau \rightarrow \infty$ in the near-horizon region. To do so, we identify the ingoing mode at the horizon $\tau \rightarrow B$ and the outgoing mode at the future infinity $\tau \rightarrow \infty$.
In the region of $\tau \rightarrow B$ $(z \rightarrow 1)$, the function $T_\pm(z)$ has the asymptotic form
\begin{eqnarray}
\lim_{z \rightarrow 1} T_\pm(z) = C_\pm (z - 1)^{\mp i (\tilde{\kappa} - \kappa)/2}.
\end{eqnarray}
Then the spinor field $\Phi$ can be decomposed into two modes
\begin{eqnarray}
\Phi_B^{(\pm)} &=& \frac{C_{\pm}}{2} \exp(i k \rho) (z - 1)^{\mp i (\tilde{\kappa} - \kappa)/2} \begin{bmatrix} i \mathbb{Y}^n_{j - 1/2} \\ \pm \mathbb{Y}^n_{j + 1/2} \end{bmatrix},
\end{eqnarray}
where the upper (lower) sign corresponds to the positive (negative) frequency modes.
Note that we need to consider the magnitudes of $\tilde{\kappa}$ and $\kappa$, which will be discussed later.
In the other region $\tau \rightarrow \infty$ $(z \rightarrow \infty)$, the function $T_\pm(z)$ can be written as
\begin{eqnarray}
\lim_{z \rightarrow \infty} T_+(z) &=& \left( C_+ \Xi_+ + C_- \frac{i m r_\mathrm{ds} +  \upsilon r_\mathrm{ds}/r_n}{i (\mu + \kappa)} \Omega_- \right) z^{-i \mu} + \left( C_+ \Omega_+ - C_- \frac{i m r_\mathrm{ds} +  \upsilon r_\mathrm{ds}/r_n}{i (\mu - \kappa)} \Xi_- \right) z^{i \mu},
\\
\lim_{z \rightarrow \infty} T_-(z) &=& \left( C_- \Omega_- + C_+ \frac{i m r_\mathrm{ds} -  \upsilon r_\mathrm{ds}/r_n}{i (\mu - \kappa)} \Xi_+ \right) z^{-i \mu} + \left( C_- \Xi_- - C_+ \frac{i m r_\mathrm{ds} -  \upsilon r_\mathrm{ds}/r_n}{i (\mu + \kappa)} \Omega_+ \right) z^{i \mu},
\end{eqnarray}
where
\begin{eqnarray}
\Xi_\pm = \frac{\Gamma\left( 1/2 \mp i (\tilde{\kappa} - \kappa) \right) \Gamma\left( \mp 2 i \mu \right)}{\Gamma\left( 1/2 \mp i (\tilde{\kappa} + \mu) \right) \Gamma\left( \mp i (\mu - \kappa) \right)}, \qquad
\Omega_\pm = \frac{\Gamma\left( 1/2 \mp i (\tilde{\kappa} - \kappa) \right) \Gamma\left( \pm 2 i \mu \right)}{\Gamma\left( 1/2 \mp i (\tilde{\kappa} - \mu) \right) \Gamma\left( \pm i (\mu + \kappa) \right)}.
\end{eqnarray}
Then the spinor field $\Phi_B$ can be decomposed into the positive frequency mode $\Phi_\infty^{(+)}$ and the negative frequency mode $\Phi_\infty^{(-)}$:
\begin{eqnarray}
\Phi_\infty^{(+)} &=& \frac{C_+ \Xi_+}{2} \exp(i k \rho) z^{-i \mu} \sqrt{\frac{2 \mu}{\mu - \kappa}} \begin{bmatrix} i \Lambda_1 \mathbb{Y}^n_{j - 1/2} \\
\Lambda_4 \mathbb{Y}^n_{j + 1/2} \end{bmatrix}
+ \frac{C_- \Omega_-}{2} \exp(i k \rho) z^{-i \mu} \sqrt{\frac{2 \mu}{\mu + \kappa}} \begin{bmatrix} i \Lambda_3 \mathbb{Y}^n_{j - 1/2} \\ - \Lambda_2 \mathbb{Y}^n_{j + 1/2} \end{bmatrix},
\\
\Phi_\infty^{(-)} &=& \frac{C_+ \Omega_+}{2} \exp(i k \rho) z^{i \mu} \sqrt{\frac{2 \mu}{\mu + \kappa}} \begin{bmatrix} i \Lambda^*_2 \mathbb{Y}^n_{j - 1/2} \\ \Lambda_3^* \mathbb{Y}^n_{j + 1/2} \end{bmatrix}
+ \frac{C_- \Xi_-}{2} \exp(i k \rho) z^{i \mu} \sqrt{\frac{2 \mu}{\mu - \kappa}} \begin{bmatrix} i \Lambda^*_4 \mathbb{Y}^n_{j - 1/2} \\ - \Lambda_1^* \mathbb{Y}^n_{j + 1/2} \end{bmatrix},
\end{eqnarray}
where
\begin{eqnarray}
&& \Lambda_1 = 1 + \frac{i m r_\mathrm{ds} - \upsilon r_\mathrm{ds}/r_n}{i (\mu - \kappa)}, \qquad
\Lambda_2 = 1 - \frac{i m r_\mathrm{ds} + \upsilon r_\mathrm{ds}/r_n}{i (\mu + \kappa)},
\nonumber\\
&& \Lambda_3 = 1 + \frac{i m r_\mathrm{ds} + \upsilon r_\mathrm{ds}/r_n}{i (\mu + \kappa)}, \qquad
\Lambda_4 = 1 - \frac{i m r_\mathrm{ds} - \upsilon r_\mathrm{ds}/r_n}{i (\mu - \kappa)},
\end{eqnarray}
Note that a necessary condition for pair production is that the parameter $\mu$ should be real; otherwise, the particle concept for spinors cannot be defined.

In the region of $B \ll 1$, the positive and negative modes depend on the relative magnitude of $\tilde{\kappa}$ and $\kappa$, and we will assume that $\tilde{\kappa} \gg \kappa$, so $\Phi_B^{(+)}$ and $\Phi_B^{(-)}$ represent the positive and negative frequency modes, respectively. To find the Bogoliubov coefficients, we impose the boundary condition that a positive frequency mode outgoes from the horizon:
\begin{eqnarray}
\Phi_B^{(-)} = 0 \quad \Rightarrow \quad C_- = 0.
\end{eqnarray}
Then, from the Bogoliubov transformation
\begin{eqnarray}
\Phi^{(+)}_B = \alpha \Phi^{(+)}_\infty + \beta^* \Phi^{(-)}_\infty,
\end{eqnarray}
we obtain the corresponding Bogoliubov coefficients
\begin{eqnarray}
\alpha = \sqrt{\frac{2 \mu}{\mu - \kappa}} \Xi_+ \quad &\Rightarrow& \quad |\alpha|^2 = \frac{\sinh(\pi \mu - \pi \kappa) \cosh(\pi \tilde{\kappa} + \pi \mu)}{\cosh(\pi \tilde{\kappa} - \pi \kappa) \sinh(2 \pi \mu)},
\\
\beta^* = \sqrt{\frac{2 \mu}{\mu + \kappa}} \Omega_+ \quad &\Rightarrow& \quad |\beta|^2 = \frac{\sinh(\pi \mu + \pi \kappa) \cosh(\pi \tilde{\kappa} - \pi \mu)}{\cosh(\pi \tilde{\kappa} - \pi \kappa) \sinh(2 \pi \mu)}.
\end{eqnarray}
They satisfy the Bogoliubov relation $|\alpha|^2 + |\beta|^2 = 1$ for fermions.

For $\tilde{\kappa} > \kappa$, introducing the Unruh and effective temperatures similarly as the scalar case~\cite{Chen:2023swn}:
\begin{equation}
T_\mathrm{U} = \frac{\kappa}{2 \pi \bar{m} r_\mathrm{ds}^2}, \qquad T_\mathrm{eff} = \sqrt{T_\mathrm{U}^2 + \frac1{(2 \pi r_\mathrm{ds})^2}} + T_\mathrm{U}, \qquad \bar{T}_\mathrm{eff} = \sqrt{T_\mathrm{U}^2 + \frac1{(2 \pi r_\mathrm{ds})^2}} - T_\mathrm{U},
\end{equation}
where $\bar{m}$ is the effective mass in the dS space
\begin{equation}
\bar{m}^2 = m^2 + \frac{(j + 1/2)^2}{r_n^2},
\end{equation}
we express the mean number in terms of thermodynamic variables as
\begin{equation}
|\beta|^2 = \frac{\left( 1 - \mathrm{e}^{-\bar{m}/\bar{T}_\mathrm{eff}} \right) \left( 1 + \mathrm{e}^{\bar{m}/T_\mathrm{eff}} \mathrm{e}^{-(k - q \Phi_\mathrm{H})/T_\mathrm{H}} \right)}{\left( \mathrm{e}^{\bar{m}/T_\mathrm{eff}} - \mathrm{e}^{-\bar{m}/\bar{T}_\mathrm{eff}} \right) \left( 1 + \mathrm{e}^{-(k - q \Phi_\mathrm{H})/T_\mathrm{H}} \right)},
\end{equation}
where the Hawking temperature and chemical potential are
\begin{equation}
T_\mathrm{H} = \frac{B}{2 \pi}, \qquad \Phi_\mathrm{H} = B \frac{Q r_\mathrm{ds}^2}{r_n^2}.
\end{equation}
The temperature $T_\mathrm{H}$ measured in the black hole coordinates is $T_\mathrm{H} = B/2 \pi \times \epsilon/r_\mathrm{ds}^2$. Note that the effective mass for scalars have $m^2 - 1/(4 r_\mathrm{ds}^2)$ and $l (l + 1)$ instead of $m^2$ and $(j + 1/2)^2$ for fermions.


\section{Inner Region of Near-Extremal Nariai Black Holes} \label{inner}

\subsection{Dirac Equation}
The timelike inner region of the near-extremal Nariai black hole is still described by the geometry of dS$_2 \times S^2$
\begin{eqnarray}
ds^2 = r_\mathrm{ds}^2 \left( - f(\rho) d\tau^2 + \frac{d \rho^2}{f(\rho)} \right) + r_n^2 \left( d\theta^2 + \sin^2\theta d\varphi^2 \right), \qquad f(\rho) = B^2 - \rho^2,
\end{eqnarray}
and the gauge (Coulomb) potential is given by
\begin{eqnarray}
A = -\frac{r_\mathrm{ds}^2 Q_n}{r_n^2} \rho d\tau = \phi(\rho) d\tau.
\end{eqnarray}
Similarly to Eq.~\eqref{out_tetrad} and replacing $e^0{}_\tau \rightarrow e^1{}_\rho$ and $e^1{}_\rho \rightarrow e^0{}_\tau$,
we define the tetrad as
\begin{eqnarray}
e^0{}_\tau = r_\mathrm{ds} \sqrt{f}, \qquad
e^1{}_\rho = \frac{r_\mathrm{ds}}{\sqrt{f}}, \qquad
e^2{}_\theta = r_n, \qquad
e^3{}_\varphi = r_n \sin\theta,
\end{eqnarray}
and write
\begin{eqnarray}
\gamma^a e_a{}^\mu \Gamma_\mu = \gamma^0 e_0{}^\tau \Gamma_\tau + \gamma^3 e_3{}^\varphi \Gamma_\varphi = \frac{\gamma^1}{4 r_\mathrm{ds} \sqrt{f}} \frac{d f}{d\rho} + \frac{\gamma^2}{2 r_n} \cot\theta.
\end{eqnarray}
Using the spinor field~\eqref{Phi_spinor},
the Dirac equation now takes the form
\begin{eqnarray}
\left( \frac{\gamma^0}{r_\mathrm{ds} \sqrt{f}} (\partial_\tau - i q \phi) + \gamma^1 \frac{\sqrt{f}}{r_\mathrm{ds}} \partial_\rho + \frac{\gamma^2}{r_n} \partial_\theta + \frac{\gamma^3}{r_n \sin\theta} \partial_\varphi + m \right) \Phi = 0.
\end{eqnarray}

\subsection{Exact Solution}
Now, we assume the positive frequency solution of the spinor field $\Phi$
\begin{eqnarray}
\Phi(\tau, \rho, \theta, \varphi) = \xi_+(\rho) \exp(-i \omega \tau) \begin{bmatrix} i \mathbb{Y}^n_{j - 1/2}(\theta, \varphi) \\ 0 \end{bmatrix}
+ \xi_-(\rho) \exp(-i \omega \tau) \begin{bmatrix} 0 \\ \mathbb{Y}^n_{j + 1/2}(\theta, \varphi) \end{bmatrix},
\end{eqnarray}
and only consider $\upsilon = - (j + 1/2)$ below.
Then the Dirac equation equals to a pair of two coupled equations
\begin{eqnarray} \label{eq_in_xi}
\left( \sqrt{f} \frac{d}{d \rho} \pm \upsilon \frac{r_\mathrm{ds}}{r_n} \right) \xi_\pm - \left( \frac{i \omega + i q \phi}{\sqrt{f}} \pm i m r_\mathrm{ds} \right) \xi_\mp = 0.
\end{eqnarray}
As in the spacelike outer region, a linear combination
\begin{eqnarray}
R_\pm(\rho) = \xi_+(\rho) \pm \xi_-(\rho) \quad \Rightarrow \quad \xi_\pm(\rho) = \frac{R_+(\rho) \pm R_-(\rho)}{2},
\end{eqnarray}
leads to
\begin{eqnarray} \label{eq_in_R}
\left( \sqrt{f} \frac{d}{d\rho} \mp \frac{i \omega + i q \phi}{\sqrt{f}} \right) R_\pm + \left( \upsilon \frac{r_\mathrm{ds}}{r_n} \pm i m r_\mathrm{ds} \right) R_\mp = 0.
\end{eqnarray}
Again, by introducing a new variable and a factor into the functions $R_\pm$
\begin{eqnarray}
z = \frac{\rho + B}{2B}, \qquad
R_\pm(z) = \Sigma^{\mp 1} \tilde{R}_\pm(z), \qquad
\Sigma = z^{- i (\tilde{\kappa} + \kappa)/2} (1 - z)^{i (\tilde{\kappa} - \kappa)/2},
\end{eqnarray}
where
\begin{eqnarray}
\tilde{\kappa} = \frac{\omega}{B}, \qquad \kappa = q Q_n \frac{r_\mathrm{ds}^2}{r_n^2},
\end{eqnarray}
we obtain
\begin{eqnarray} \label{eq_in_tR}
\Sigma^{\mp 2} \sqrt{z (1 - z)} \, \frac{d\tilde{R}_\pm}{dz} + \left( \upsilon \frac{r_\mathrm{ds}}{r_n} \pm i m r_\mathrm{ds} \right) \tilde{R}_\mp = 0.
\end{eqnarray}
Finally, the solutions of Eq.~\eqref{eq_in_tR} are given by the hypergeometric functions
\begin{eqnarray}
\tilde{R}_\pm &=& C_\pm F\left( \pm i (\mu + \kappa), \mp i (\mu - \kappa); \frac{1}{2} \pm i (\tilde{\kappa} + \kappa); z \right)
\nonumber \\
&+& D_\pm \Sigma^{\pm 2} \sqrt{z (1 - z)} F\left( 1 \mp i (\mu + \kappa), 1 \pm i (\mu - \kappa); \frac{3}{2} \mp i (\tilde{\kappa} + \kappa); z \right),
\end{eqnarray}
where
\begin{eqnarray}
\mu^2 = \kappa^2 + m^2 r_\mathrm{ds}^2 + \upsilon^2 \frac{r_\mathrm{ds}^2}{r_n^2}, \qquad D_\pm = - C_\mp \frac{\upsilon r_\mathrm{ds}/r_n \pm i m r_\mathrm{ds}}{1/2 \mp i (\tilde{\kappa} + \kappa)}.
\end{eqnarray}

\subsection{Near Horizon Behaviors}
As $z \rightarrow 0$ $(\rho \rightarrow -B)$, the function $R_\pm(z)$ has the asymptotic form
\begin{eqnarray}
\lim_{z \rightarrow 0} R_\pm(z) = C_\pm z^{\pm i (\tilde{\kappa} + \kappa)/2}.
\end{eqnarray}
Then, the spinor field $\Phi$ can be decomposed into the outgoing (ingoing) modes $\Phi_{-B}^{(\pm)}$ as
\begin{eqnarray}
\Phi_{-B}^{(\pm)} &=& \frac{C_{\pm}}{2} \exp(- i \omega \tau) z^{\pm i (\tilde{\kappa} + \kappa)/2} \begin{bmatrix} i \mathbb{Y}^n_{j - 1/2} \\ \pm \mathbb{Y}^n_{j + 1/2} \end{bmatrix}.
\end{eqnarray}
As $z \rightarrow 1$ $(\rho \rightarrow B)$, the function $R_\pm(z)$ can be written as
\begin{eqnarray}
\lim_{z \rightarrow 1} R_\pm(z) = \left( C_\pm \Xi_\pm - C_\mp \Omega_\pm \right) (1 - z)^{\mp i (\tilde{\kappa} - \kappa)/2},
\end{eqnarray}
where
\begin{eqnarray}
\Xi_\pm = \frac{\Gamma\left( 1/2 \pm  i (\tilde{\kappa} + \kappa) \right) \Gamma\left( 1/2 \pm i (\tilde{\kappa} - \kappa) \right)}{\Gamma\left( 1/2 \pm i (\tilde{\kappa} + \mu) \right) \Gamma\left( 1/2 \pm i (\tilde{\kappa} - \mu) \right)},
\quad
\Omega_\pm = \left( \upsilon \frac{r_\mathrm{ds}}{r_n} \pm i m r_\mathrm{ds} \right) \frac{\Gamma\left( 1/2 \mp i (\tilde{\kappa} + \kappa) \right) \Gamma\left( 1/2 \pm i (\tilde{\kappa} - \kappa) \right)}{\Gamma\left( 1 \mp i (\mu + \kappa) \right) \Gamma\left( 1 \pm i (\mu - \kappa) \right)}.
\end{eqnarray}
Then the spinor field $\Phi$ can be decomposed into two modes
\begin{eqnarray}
\Phi_B^{(+)} &=& \left( \frac{C_+ \Xi_+}{2} - \frac{C_- \Omega_+}{2} \right) \exp(- i \omega \tau) (1 - z)^{- i (\tilde{\kappa} - \kappa)/2} \begin{bmatrix} i \mathbb{Y}^n_{j - 1/2} \\ \mathbb{Y}^n_{j + 1/2} \end{bmatrix},
\\
\Phi_B^{(-)} &=& \left( \frac{C_- \Xi_-}{2} - \frac{C_+ \Omega_-}{2} \right) \exp(- i \omega \tau) (1 - z)^{i (\tilde{\kappa}-\kappa)/2} \begin{bmatrix} i \mathbb{Y}^n_{j - 1/2} \\ - \mathbb{Y}^n_{j + 1/2} \end{bmatrix}.
\end{eqnarray}

In the near-extremal limit $B \rightarrow 0$, we assume that $\tilde{\kappa} \gg \kappa$ and impose the boundary condition
\begin{eqnarray}
\Phi_B^{(-)} = 0 \quad \Rightarrow \quad C_- = C_+ \frac{\Omega_-}{\Xi_-}.
\end{eqnarray}
By applying the Bogoliubov transformation
\begin{eqnarray}
\Phi^{(+)}_{-B} + \alpha \Phi^{(-)}_{-B} = \beta^* \Phi^{(+)}_{B},
\end{eqnarray}
we obtain the corresponding Bogoliubov coefficients
\begin{eqnarray}
\alpha = \frac{\Omega_-}{\Xi_-} \quad &\Rightarrow& \quad |\alpha|^2 = \frac{\sinh(\pi \mu + \pi \kappa) \sinh(\pi \mu - \pi \kappa)}{\cosh(\pi \tilde{\kappa} + \pi \mu) \cosh(\pi \tilde{\kappa} - \pi\mu)},
\\
\beta^* = \frac{1}{\Xi_-} \quad &\Rightarrow& \quad |\beta|^2 = \frac{\cosh(\pi \tilde{\kappa} + \pi \kappa) \cosh(\pi \tilde{\kappa} - \pi \kappa)}{\cosh(\pi \tilde{\kappa} + \pi\mu) \cosh(\pi \tilde{\kappa} - \pi \mu)}. \label{inner_emission}
\end{eqnarray}
They satisfy the Bogoliubov relation. Remarkably, as $\tilde{\kappa} \rightarrow \infty$ in the exact Nariai limit, the mean number approaches the unity that saturates the bound from the Pauli blocking due to the exclusion principle. That is, as two spherical shells of the cosmological constant and black hole horizon collapse, the mean number $|\beta|^2 = 1$ while $|\alpha|^2 = 0$.

%

\section{Fermion Production in ${\rm (A)dS_2 \times S^2}$} \label{(A)dS_2}
We have shown that as the near-extremality disappears ($B \rightarrow 0$) and Nariai black holes become an exact limit, the fermion production saturates the bound from the Pauli blocking between the black hole horizon and cosmological horizon, i.e, $|\beta|^2 \rightarrow 1$.
In the exact Nariai limit, the outer region reduces to ${\rm dS_2 \times S^2}$.

With a rescaling $\tau = r_{\rm ds} \mathrm{e}^{t/r_{\rm ds}}, \rho = r_\mathrm{ds}^{-2} x$, the exact Nariai metric (\ref{Nariai-metric}) $(B = 0)$ can be written as
\begin{eqnarray}
ds^2 = - dt^2 + \mathrm{e}^{2 t/r_{\rm ds}} dx^2 + r_n^2 d \Omega_2^2,
\end{eqnarray}
which is ${\rm dS_2 \times S^2}$ in the planar coordinates. In other word, the outer region of Nariai black hole corresponds to the dS space in the planar coordinates. One may thus expect that in the exact Nariai limit $(\tilde{\kappa} = \infty)$ the mean number
\begin{eqnarray}
|\beta|^2 = \frac{\sinh(\pi \mu + \pi \kappa)}{\mathrm{e}^{\pi \mu - \pi \kappa} \sinh(2 \pi \mu)},
\end{eqnarray}
has the same form as that of fermions in the planar coordinates of ${\rm dS_2}$~\cite{Chen:2025xrv}.

On the other hand, the extremal RN black hole in dS space has the near-horizon geometry~\cite{Chen:2020mqs}
\begin{eqnarray}
ds^2 = - \frac{\rho^2}{r_{\rm ads}^2} d\tau^2 + \frac{r_{\rm ads}^2}{\rho^2} d\rho^2 + r_n^2 d\Omega_2^2,
\end{eqnarray}
which, after rescaling the coordinate $\rho = r_{\rm ads} \mathrm{e}^{x/r_{\rm ads}}$, can be written as the product ${\rm AdS_2 \times S^2}$ in the planar coordinates
\begin{eqnarray}
ds^2 = - \mathrm{e}^{2 x/r_{\rm ads}} d\tau^2 + dx^2 + r_n^2 d\Omega_2^2.
\end{eqnarray}
According to the reciprocal relation $\mathcal{N}_\mathrm{ds}(R) \mathcal{N}_\mathrm{ads}(R) = 1$~\cite{Chen:2025xrv}, the mean number of spontaneously produced fermion (spinor) pairs becomes
\begin{eqnarray}\label{ex-RN}
{\cal N}^{(\rm sp)}_{\rm eRN} = \frac{\sinh(2 \pi \mu (r_{\rm ads}))}{\mathrm{e}^{\pi \kappa - \pi \mu(r_{\rm ads})} \sinh(\pi \kappa + \pi \mu(r_{\rm ads}))},
\end{eqnarray}
where
\begin{eqnarray}
\mu (r_{\rm ads}) = \sqrt{\kappa_{\rm ads}^2 - \bar{m}^2 r_{\rm ads}^2}, \qquad \kappa_{\rm ads} = q Q_n \frac{r_{\rm ads}^2}{r_n^2}.
\end{eqnarray}
The result~(\ref{ex-RN}) agrees with the mean number for fermion production by a uniform electric field in ${\rm AdS_2}$.

It should be noted that the spontaneous production of fermion pairs is blocked by the Pauli exclusion principle both in extremal charged Nariai and RN black holes in an asymptotic dS space. Assuming the reciprocal relation between the near-extremal Nariai black hole and the near-extremal RN black hole, the mean number for fermion production in near-extremal RN black hole would become
\begin{eqnarray}
{\cal N}^{(\rm sp)}_{\rm neRN} = \frac{\cosh(\pi \tilde{\kappa}_{\rm ads} + \pi\mu (r_{\rm ads})) \cosh(\pi \tilde{\kappa}_{\rm ads} - \pi \mu (r_{\rm ads}) )}{\cosh(\pi \tilde{\kappa}_{\rm ads} + \pi \kappa_{\rm ads}) \cosh(\pi \tilde{\kappa}_{\rm ads} - \pi \kappa_{\rm ads})},
\end{eqnarray}
where $\tilde{\kappa}_{\rm ads} = \omega / B$.

The spontaneous fermion production can be compared with the spinless boson production. In the inner region of near-extremal Nariai black hole, the mean number of bosons for $\tilde{\kappa} \gg 1$
\begin{eqnarray}
{\cal N}^{(\rm sc)}_{\rm in} = \frac{\mathrm{e}^{2 \pi \tilde{\kappa}}}{\cosh (\pi \mu + \pi \kappa) \cosh (\pi \mu - \pi \kappa)}
\end{eqnarray}
exponentially explodes. The catastrophic boson production is a strong contrast to fermion production with saturation of unity. The mean numbers of bosons from the Nariai black hole and extremal RN black hole have the same forms as in ${\rm dS}_2$ and ${\rm AdS}_2$~\cite{Cai:2014qba} and also satisfy the reciprocal relation~\cite{Chen:2025xrv}.

\section{Comparison of Emission of Spinors and Scalars} \label{spinor_scalar}
\begin{figure}[ht]
\includegraphics[scale=0.7, angle=0]{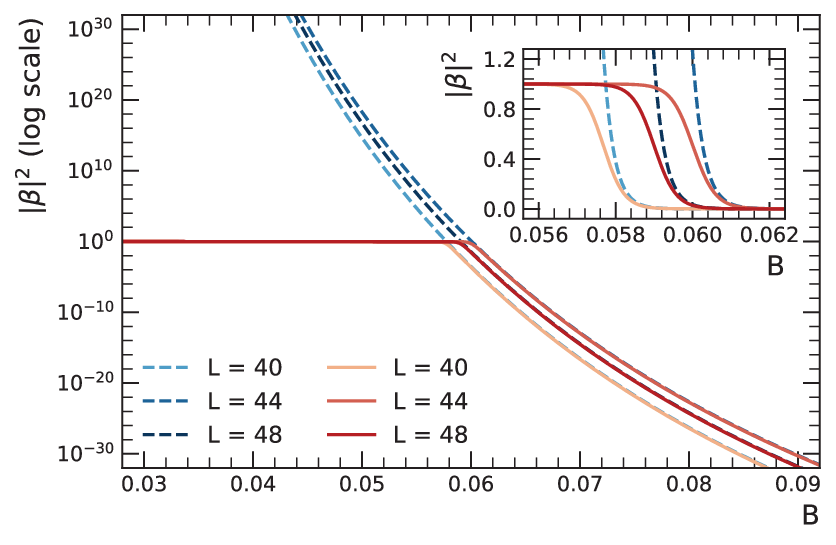}
\caption{Comparison of the emission of fermions and scalars in near-extremal Nariai black holes. Dotted lines for scalars and solid lines for fermions. Figures are drawn for $m = q =1, Q_n =10, \omega = 2$ and the s-wave $l = 0$.}
\label{Pauli_block}
\end{figure}

The mean number~\eqref{inner_emission} of fermions emitted in the inner region may be written in terms of the Hawking temperature for near-extremal Nariai black holes, the Unruh temperature for charge acceleration, and the Gibbons-Hawking temperature for ${\rm dS}_2$ space as
\begin{eqnarray}
N_\mathrm{in}^{(\mathrm{sp})} = \frac{\cosh\Bigl( \frac{\omega}{2 T_H} + \frac{\bar{m} T_U}{2 T_C^2} \Bigr) \cosh\Bigl( \frac{\omega}{2 T_H} - \frac{\bar{m} T_U}{2 T_C^2} \Bigr)}{\cosh\Bigl( \frac{\omega}{2 T_H} + \frac{\bar{m} \sqrt{T_C^2 + T_U^2}}{2 T_C^2} \Bigr) \cosh\Bigl( \frac{\omega}{2 T_H} - \frac{\bar{m} \sqrt{T_C^2 + T_U^2}}{2 T_C^2} \Bigr)}.
\end{eqnarray}
The leading terms up to $\mathrm{e}^{-\omega/T_H}$ are approximately given by
\begin{eqnarray}
N_{\mathrm{in}}^{(\mathrm{sp})} \approx 1 - 4 \, \mathrm{e}^{- \frac{\omega}{T_H}} \sinh\Bigl( \frac{\bar{m} T_{\mathrm{eff}}}{2 T_C^2} \Bigr)
\sinh\Bigl( \frac{\bar{m} \bar{T}_{\mathrm{eff}}}{2 T_C^2} \Bigr).
\end{eqnarray}
Thus, the spinor emission rapidly reaches the limit $(|\beta|^2 = 1)$ due to the Pauli-blocking when $B$ $(T_H)$ approaches to zero,
as shown in Fig.~\ref{Pauli_block}.

It is interesting to compare the spinor emission with that of the scalar emission~\cite{Chen:2023swn} in the region of $B \ll 1$, which may be written as
\begin{eqnarray}
N_{\mathrm{in}}^{(\mathrm{sc})} = \frac{\mathrm{e}^{- \frac{q \Phi_H}{T_H}} \sinh\Bigl( \frac{\omega + q \Phi_H}{2 T_H} \Bigr) \sinh\Bigl( \frac{\omega - q \Phi_H}{2 T_H} \Bigr)}{\mathrm{e}^{- \frac{\bar{m} T_U}{T_C^2}} \cosh\Bigl( \frac{\bar{m} T_{\mathrm{eff}}}{2 T_C^2} \Bigr) \cosh\Bigl( \frac{\bar{m} \bar{T}_{\mathrm{eff}}}{2 T_C^2} \Bigr)}.
\end{eqnarray}
Indeed, there is a catastrophic emission of the form $\mathrm{e}^{(\omega - q \Phi_H)/T_H}$ provided that $\omega > q \Phi_H$, which is the opposite of superradiance~\cite{Brito:2015oca}, which is shown in Fig.~\ref{Pauli_block}.

\section{Conclusion} \label{conclusion}
The pair production of scalars and spinors in background fields is determined by the Bogoliubov relations in the in-out formalism that obey the spin statistics. The Bogoliubov coefficients satisfy the relations
\begin{eqnarray}
|\alpha|^2 \mp |\beta|^2 = 1,
\end{eqnarray}
where the upper (lower) sign is boson (fermion). Thus, the mean number for boson production can have an amplification greater than one while the mean number of fermion production cannot be greater than one. Near-extremal Nariai black holes have a timelike region between two horizons while near-extremal RN black holes have a spacelike region between two horizons. It has been shown that the emission of charged bosons from (near-)extremal black holes exhibits the characteristic of the Schwinger formula in electric profiles in the Minkowski spacetime, which is not greater than one and whose leading Boltzmann factor is given by non-zero instanton action~\cite{Chen:2012zn, Chen:2020mqs, Cai:2020trh, Chen:2023swn}, and even the rotation of black holes does change this feature~\cite{Chen:2016caa, Chen:2017mnm, Chen:2021jwy, Chen:2024ctf}. This no-bosonic amplification seems to contradict to our intuition. However, we have shown in~\cite{Chen:2023swn} that near-extremal Nariai black holes emit charged bosons catastrophically, similar to two narrow spherical shells with a high potential difference.

In this paper we have studied the spontaneous production of spin-1/2 fermions from near-extremal Nariai black holes. The scattering of fermions contrasts with that of bosons in black hole background. Bosons can have superradiance modes, in which the reflected flux becomes greater than the incident flux due to pair production~\cite{2020LNP971B}. Hence it is interesting to study the emission of charged fermions from Nariai black holes. We have found that the mean number of fermions in the near-extremal Nariai black holes still satisfies the Bogoliubov relation but saturates the bound from the Pauli blocking due to the exclusion principle.

From the view of quantum field theory, spontaneous pair or particle production is the result of one-loop effective actions, as explicitly shown by Heisenberg-Euler~\cite{Heisenberg1936Folgerungen} and Schwinger~\cite{Schwinger:1951nm} in the one-loop QED action for charged scalars or spinors in a constant electromagnetic field. A constant electric field produces pairs of bosons or fermions while a magnetic field is stable against pair production, whose one-loop QED action does not have poles that give an imaginary part due to pair production. It is thus uttermost interesting to investigate fermion systems that saturate the bound from the Pauli blocking.

\acknowledgments
C.M.C. and S.P.K. thank the participants of the joint program [APCTP-2025-J01] held at APCTP, Pohang, Korea for fruitful discussions. The work of C.M.C. was supported by the National Science Council of the R.O.C. (Taiwan) under the grant NSTC 114-2112-M-008-010.

\bibliography{ref_Nariai_spinor.bib}

\end{document}